1# Biological Sciences - Neuroscience

# Loss of AP3 function affects spontaneous and evoked release at hippocampal mossy fiber synapses

Anita Scheuber*¶, Rachel Rudge†¶, Lydia Danglot†¶, Graca Raposo‡, Thomas Binz§, Jean-Christophe Poncer* and Thierry Galli†.

\* *INSERM, Université Pierre et Marie Curie-Paris 6, UMR739, "Cortex & Epilepsy", Paris, F-75013, France.*

†*"Membrane Traffic in Neuronal and Epithelial Morphogenesis", INSERM Avenir Team, Paris, F-75005 France; Institut Jacques Monod, CNRS UMR7592, Universitities Paris 6 & Paris 7, Paris, F-75005 France.*

‡ *"Structure and Membrane Compartments", CNRS UMR 144, Institut Curie, Paris, F-75005, France.*

§ *Institute of Biochemistry, Medical School Hannover, D-30625 Hannover, Germany.*

¶*These authors contributed equally to this work and are listed in reverse alphabetical order.*

Corresponding author: Thierry Galli, Institut Jacques Monod, 2, place Jussieu, F-75251 Paris Cedex 05, France, mailto:thierry@tgalli.net, tel: +33 144 278 211, fax: +33 144 278 210, http://thierry.galli.free.fr/**Abbreviations footnote:** Impact of loss of AP3 in Neurotransmitter Release

2**Abstract**

**Synaptic vesicle exocytosis mediating neurotransmitter release occurs spontaneously at low intraterminal calcium concentrations and is stimulated by a rise in intracellular calcium. Exocytosis is compensated for by the reformation of vesicles at plasma membrane and endosomes. Although the adaptor complex AP3 was proposed to be involved in the formation of synaptic vesicles from endosomes, whether its function has an indirect effect on exocytosis remains unknown. Using *mocha* mice, which are deficient in functional AP3, we identify an AP3-dependent, tetanus neurotoxin-resistant, asynchronous release that can be evoked at hippocampal mossy fiber synapses. Presynaptic targeting of the tetanus neurotoxin-resistant vesicle SNARE TI-VAMP is lost in *mocha* hippocampal mossy fiber terminals while the localization of synaptobrevin 2 is unaffected. In addition, quantal release in *mocha* cultures is more frequent, and more sensitive to sucrose. We conclude that the lack of AP3 results in more constitutive secretion and the loss of an asynchronous evoked release component, suggesting an important function of AP3 in regulating SV exocytosis at mossy fiber terminals.**

## Introduction

The release of neurotransmitter at the synapse requires the fusion and recycling of synaptic vesicles (SVs). The fusion of SVs with the plasma membrane depends on the formation of SNARE complexes between vesicle SNAREs and plasma membrane SNAREs, as demonstrated by the striking sensitivity of neurotransmitter release to clostiridial neurotoxins particularly Tetanus Neurotoxin (TeNT, for a review see (1)). Several models of the recycling of synaptic vesicles have been proposed: endosomal recycling, SV budding from the plasma membrane (2), kiss-and-run and kiss-and-stay (for a review see (3)). Endosomal recycling involves the molecular coat AP3 as suggested from experiments in neuroendocrine cells (4) but the importance of AP3 in neurotransmitter release is still unclear. AP3 is composed of four subunits and two different AP3 complexes are expressed in brain: the ubiquitous AP3A, composed of the δ, σ3, β3A and μ3A subunits, and the neuronal specific AP3B, composed of the δ, σ3, β3B and μ3B subunits (5, 6). *Mocha* mice are deficient for the δ subunit and therefore lack both AP3A and AP3B complexes. These mice have neurological disorders including hyperactivity and spontaneous seizures. In this paper, we set out to understand the importance of AP3 function in neurotransmitter release by characterizing basal and evoked release neurotransmitter release in *mocha* mice.

## Results

*Asynchronous release evoked at mossy fiber terminals is lost in mocha mice*

AP3 is particularly concentrated in the hilus and CA3 region of the hippocampus in heterozygous control (+/-) mice (Fig. S1a). Therefore, in order to assess the role of AP3 in neurotransmitter release, we compared synaptic transmission at mossy fiber (MF)-CA3 synapses in organotypic hippocampal cultures from *mocha* (-/-) and heterozygous control (+/-) littermates (Fig. 1a). We first examined $Ca^{2+}$-dependent release evoked by MF stimulation. A stimulating electrode placed at the hilar border of the granule cell layer reliably evoked large post-synaptic currents (PSCs) that were specifically suppressed by the group 2 metabotropic glutamate receptor agonist DCG-IV (Fig. 1b; (7)). The amplitude of the PSCs gradually increased with stimulation intensities ranging from 15 to 600 V.μs. The correlation between the average charge of the PSCs and the stimulation intensity was not significantly different in cultures prepared from control vs. *mocha* mice (Fig. 1c). Cleavage of the SV SNARE synaptobrevin 2 (Syb2) by preincubation of the cultures with tetanus neurotoxin (TeNT) for >72h caused a dramatic reduction in transmitter release evoked by MF stimulation. In cultures prepared from control mice, stimuli up to 2000 V.μs evoked small, unreliable PSCs, reminiscent of evoked transmission in cultured hippocampal neurons from Syb2 knockout mice (8). Release at stimulated synapses was usually asynchronous and occurred within ~200 ms after stimulation with an average probability of 0.26±0.05 at the highest stimulation intensities (n=7). In contrast, no PSC could be evoked by MF stimulation in cultures prepared from *mocha* littermate mice.

*Loss of presynaptic TI-VAMP in mossy fiber terminals*

Our observations may reflect the contribution of a TeNT-insensitive, AP3-dependent pathway of transmitter release at MF terminals. We have previously shown that TI-VAMP (i) is present at a high level in SVs at MF terminals (9), (ii) interacts with the δ subunit of the AP3 complex and (iii) is mistargeted in *mocha* fibroblasts (10). TI-VAMP is therefore the best candidate v-SNARE to support TeNT-resistant vesicle exocytosis at MF terminals. Since TI-VAMP is not expressed at Schaffer collateral terminals onto rat CA1 pyramidal cells ((9), Fig. S1a), we anticipated this form of exocytosis may not be observed at these synapses. Consistent with this prediction, evoked release was entirely impaired by TeNT at Schaffer collateral synapses onto CA1 pyramidal cells in control mice (Fig. 1a,b). We therefore compared TI-VAMP localization in hippocampal sections from control and *mocha* mice. In the CA3 and dentate gyrus areas of control mice, TI-VAMP was localized in the MF terminals, as demonstrated by its colocalization with the presynaptic marker synaptophysin (Syp, Fig. 2a, S1a). This was further confirmed in cultured granule cells because we found that 67.20+/-6.65% of Syp positive punctae were also TI-VAMP positive (Fig. S2a). However in *mocha* sections, TI-VAMP labeling in MFs was completely lost (Fig. 2a, S1b). Thus, AP3 is required for the presynaptic targeting of TI-VAMP to MF terminals. In contrast, Syb2 localization was unchanged in brain sections from *mocha* mice as compared to control (Fig. 2b) suggesting that the presynaptic targeting of Syb2 is independent of AP3. Other SV proteins, including Rab3a, synaptotagmin 1, showed a presynaptic targeting similar in control and *mocha* mice (Fig. S1b, and our unpublished observations). Together, these results suggest that a form of TeNT-insensitive, AP3-dependent, evoked release exists at MF-CA3 synapses, which likely involves TI-VAMP as a v-SNARE.

*TI-VAMP is blocked in the cell body of mocha neurons*

We then examined the subcellular localization of TI-VAMP in *mocha* hippocampal granule cells from which mossy fibers originate. We found that TI-VAMP accumulated in the cell bodies of granule cells as shown by colocalization with VAMP4, a vesicular SNARE located in early endosomes (11, 12) and the TGN (13) (Fig. 2d). However, the AP-3 dependent sorting of TI-VAMP in the perinuclear VAMP4-positive compartment may not be specific to granule cells since a strong colocalization of TI-VAMP and VAMP4 was also found in the *mocha* CA3 pyramidal cells (Fig. 2c). The accumulation of TI-VAMP in cell bodies in *mocha* neurons was also observed in cultured hippocampal pyramidal neurons by immunolabeling (Fig. S2b) as well as immunogold labeling in ultrathin cryosections analyzed by electron microscopy (Fig. S2c). TI-VAMP labeling in *mocha* neurons was found restricted to the cytosolic side of Golgi cisternea whereas in control pyramidal neurons TI-VAMP labeled vesicles and the plasma membrane, consistent with its function as a secretory v-SNARE in control but not *mocha* neurons. Therefore, an AP-3 dependent sorting of TI-VAMP at the level of a VAMP4-positive perinuclear compartment is required for the proper targeting of TI-VAMP.

*Increased basal release in mocha and BFA-treated mossy fiber terminals*

We then asked whether the lack of AP3 and TI-VAMP may impact $Ca^{2+}$-independent, constitutive release at MF terminals from *mocha* mice. Miniature EPSCs (mEPSCs) were recorded from CA3 pyramidal cells in slice cultures prepared from control and *mocha* littermates. In cultures from *mocha* mice, the frequency of mEPSCs was ~2-fold higher than in control cultures ($3.22\pm0.55$ vs. $1.66 \pm 0.72$ Hz, p<0.03, Fig. 3a). However, their mean amplitude was unchanged ($19.44 \pm 1.73$ vs. $18.97 \pm 3.32$ pA, p=0.91) as were their rise time (10-90% of peak, $1.59 \pm 0.09$ vs. $1.45 \pm 0.19$ ms, p=0.32) and decay time constant ($3.86 \pm 0.12$ vs. $3.37 \pm 0.24$ ms, p=0.14; Fig. 3b,c). These results suggest the lack of AP3 did not alter the rate of fusion pore opening or the number of postsynaptic receptors at excitatory synapses onto CA3 pyramidal cells. In addition, the higher mEPSC frequency in *mocha* cultures was not affected by the NMDA receptor antagonist APV (our unpublished observations), suggesting it did not reflect a greater activation of presynaptic NMDA receptors (14, 15) due to the lack of Zn release from MF terminals in *mocha* mice (16, 17). After 72h incubation with TeNT, mEPSC frequency in control cultures was reduced by ~84 % (to $0.27 \pm 0.04$ Hz, n=9; Fig. 3*d*), with no apparent change in their mean amplitude ($19.18 \pm 0.99$ vs. $18.97 \pm 3.32$ pA, p=0.89), again consistent with observations in cultured hippocampal neurons from Syb2 knockout mice (8). In contrast, quantal release in *mocha* cultures was more resistant to TeNT and was reduced in frequency by only ~44 % (to $1.79 \pm 0.56$ Hz, n=8). These observations suggest the absence of AP3 not only increases Ca-independent quantal release but also reduces the effect of TeNT. Apart from TI-VAMP and Syb2, no other v-SNARE protein is known to be present at excitatory synapses on CA3 pyramidal cells. Therefore, we analyzed the penetration and cleavage efficiency of TeNT in the slice cultures. We labeled TeNT-*treated* slice cultures with DAPI and antibodies against TeNT and SNAP-25. We found that TeNT penetrated throughout *mocha* (Fig. 3e) as well as control explants (our unpublished observations). Furthermore, although the vast majority of Syb2 was cleaved in control and *mocha* explants after 72h incubation with TeNT, a small fraction of Syb2 was still detected by western blotting. Quantification of the blots revealed twice as much TeNT-resistant Syb2 protein in *mocha* cultures as compared to control (Fig. 3f). Treatment with TeNT drastically reduced Syb2 labeling, but the remaining signal corresponded largely to Syb2 present at synapses, as revealed by colocalization with synaptophysin (Fig. 3g) thus the numerous TeNT-resistant mEPSCs observed in *mocha* are likely to be mediated by a presynaptic pool of Syb2 that resisted the treatment with TeNT.

Newly formed vesicles derived from endosomes disappear with a half-life of ~36 min after treatment with the fungal drug Brefeldin A (BFA) which specifically targets the AP3-dependent pathway in PC12 cells (4, 18-20). In order to test the effect of acute inactivation of the AP3 pathway, we therefore incubated control hippocampal slices with BFA for ~2-4 hours. In BFA-treated slices, mEPSC frequency was increased by ~127 % as compared to control (2.59 ± 0.33 vs. 1.14 ± 0.25 Hz, n=14 and 12 cells, respectively, p<0.005, Fig. S4a-c). The ratio of synaptic TI-VAMP/synaptic Syp measured by immunolabeling was not significantly altered by BFA neither in the hilus nor in the st. lucidum (Fig. S4d). This is consistent with the fact that longer times may be required to clear TI-VAMP from MF terminals and demonstrate that the *mocha* phenotype can be reproduced by acute pharmacological inactivation of AP3-dependent SV formation.

*Increased sensitivity of release to osmotic stimulation in mocha MF terminals*

Syb2 is resistant to TeNT (21) in SNARE complexes (21) that may be clamped by complexin and synaptotagmin before calcium rise (22). TeNT-resistant Syb2 was associated with release-competent SVs (23, 24). Our previous results could suggest that the lack of AP3 and TI-VAMP in MF SVs may thus increase the capacity of Syb2 to form TeNT-resistant clamped SNARE complexes thereby enhancing the probability of calcium-independent fusion at *mocha* MF terminals. In order to test this hypothesis, we examined the rate of release induced by focal application of a hypertonic solution because previous studies showed that hypertonic solution specifically recruits readily releasable quanta at hippocampal synapses (25) and stimulates secretion in a calcium independent manner. We thus compared the effects of focal applications of a 0.5 M sucrose solution through a patch pipette positioned in st. lucidum ~25 µm away from the somata of the recorded pyramidal cells. Since effective sucrose concentration at release sites may be difficult to control in slice cultures, we used varying injection pressures to compare the sucrose sensitivity of release in *mocha* vs. control cultures. In cells recorded from control cultures, application of sucrose with low pressure (0.25 psi) caused a ~2-fold increase in mEPSC frequency, whereas at a higher pressure (1.5 psi), a further ~7-fold increase was observed (Fig. 4a, b, c). The sensitivity of the release rate to sucrose was significantly increased in recordings from *mocha* cultures: even low pressure application of sucrose caused a ~9-fold increase in mEPSC frequency, which was further enhanced by another ~35 % at high pressure (Fig. 4a, b, c). Interestingly however, recruitment of readily releasable vesicles by sucrose was disrupted in both control and *mocha* cultures by prior incubation with TeNT, and application of hypertonic solution failed to produce an increase in mEPSC frequency even at high pressure (Fig. 4a, b). Taken together, these results suggest that an AP3-dependent mechanism decreases the sucrose sensitivity of constitutive secretion.

**Discussion**

*Mocha* mice are deficient for AP3δ subunit and therefore lack both ubiquitous AP3A and neuronal AP3B complexes. Here, we have shown that presynaptic TI-VAMP, a well-established AP-3 δ cargo (10, 26), is lost in *mocha* CA3 MFs. Other AP-3 cargos including the zinc transporter ZnT-3 (16, 17), the chloride channel ClC-3 (27), and the phosphatidylinositol-4-kinase type IIα (28) are mislocalized in *mocha* CA3 MFs. A previous study in ZnT3 knock out mice showed the lack of vesicular zinc in MFs does not significantly affect the MF-associated excitability of CA3 pyramidal cells (29). In addition, we observed that the higher mEPSC frequency in *mocha* cultures was not affected by blocking NMDA receptors, further suggesting that the lack of Zn release from MF terminals in *mocha* mice does not explain the phenotype we observed. Similarly, the lack of MF ClC-3 is unlikely to explain the increased quantal release from *mocha* MF terminals mice. The loss of ClC-3 rather affects acidification of SVs resulting in a slight reduction of quantal size (30). Finally, recent data suggested AP3 may regulate the volume of large dense-core vesicles in chromaffin cells (31). This however is unlikely to apply to MF terminals since i) we did not observe any change in quantal size in *mocha* cultures and ii) the pathway of SV reformation largely differs from that of secretory granules, the latter maturing via the removal of material from immature secretory granules. Treatment of hippocampal slice cultures with TeNT revealed an asynchronous component of secretion evoked in CA3 pyramidal cells by single stimulation of mossy fibers. This asynchronous release is unlikely to be mediated by the low amount of TeNT-resistant Syb2 since it was not observed in *mocha* cultures which showed more TeNT-resistant Syb2. Furthermore, this asynchronous component was observed at MF-CA3 synapses but not at CA1-CA3 synapses, where TI-VAMP is not expressed. Since no other TeNT-resistant v-SNARE was ever detected at MF terminals, we suggest TI-VAMP likely mediates the asynchronous evoked release unraveled in our experiments. In conclusion, although we cannot exclude that other AP3 cargos lost in *mocha* MF terminals may participate to the *mocha* phenotype, the loss of TI-VAMP seems most likely to explain the perturbation of evoked SV release described in the present study.

Asynchronous release was not observed in TeNT-untreated explants, suggesting inactivation of Syb2 may be required for the expression of this asynchronous evoked release. This observation strongly suggests TI-VAMP and Syb2 may both be present on the same rather than distinct SVs. Consistent with this scenario, Syb2 was shown to have a higher rate of SNARE complex assembly than TI-VAMP both *in vitro* and *in vivo* (10), predicting that evoked release mediated by TI-VAMP would be detected only after cleavage of Syb2. In addition, cleavage of Syb2 by TeNT was shown to modify the coupling of intracellular calcium and release-competent vesicles (32) suggesting that removal of TeNT-sensitive v-SNAREs allows for the expression of a secretory machinery that may be hard to observe otherwise. Interestingly, $Sr^+$ preferentially stimulates asynchronous release (33) and different synaptotagmin isoforms show different sensitivities to $Ca^{2+}$ and $Sr^+$ (34). For instance, Synaptotagmin 1 and 7 have different sensitivities to calcium (35), the latter interacting with TI-VAMP in fibroblasts (36) and showing biochemical properties suitable for a $Ca^{2+}$ sensor for asynchronous release (35, 37, 38). Therefore, we speculate that MF SVs may be equipped with two distinct fusion machineries for exocytosis, one depending on Syb2 and the other on TI-VAMP as a v-SNARE, which may be recruited in different conditions.

We have shown a higher basal, calcium-independent release in *mocha* cultures, with no apparent change in quantal size or current kinetics. This observation strongly argues for a presynaptic difference between control and *mocha* MF-CA3 synapses. This likely not due to a greater density of MF terminals onto *mocha* CA3 pyramidal cells because of axonal sprouting induced in organotypic cultures (39) or a greater contribution of recurrent CA3-CA3 synapses to miniature EPSCs in mocha cultures. Indeed, we found an equal Syp staining in control and *mocha* brain sections and the increased quantal release was also observed upon pharmacological disruption of AP3 by BFA in acute hippocampal slices. In addition, quantal release evoked by hyperosmotic solution directly applied onto MF terminals was also increased in *mocha* cultures. Therefore, the most likely explanation is that more vesicles are ready-to-fuse due to the absence of AP3 and possibly TI-VAMP in *mocha* vs. control MF terminals. Since BFA specifically impairs the AP3-dependent formation of new vesicles from endosomes our results suggest that AP3-dependent, newly-formed vesicles have a reduced capacity to fuse. How AP3-dependent sorting may decrease basal release remains to be explored.

In conclusion, our data suggest that the molecular mechanism of transmitter release at MF terminals reaches a high degree of complexity with at least two exocytic SV-SNAREs: Syb2 and TI-VAMP. AP3 function is important for both constitutive as well as evoked release, raising the possibility that specific forms of synaptic plasticity might occur at terminals expressing AP3 cargos like TI-VAMP (9).

## Materials and methods

### Animals

Heterozygous *mocha* mice were originally obtained from M. Robinson (CIMR, Cambridge) and then bred in-house. The experiments were carried out in accordance with the European Communities Council Directive of 24 November 1986 (86/609/EEC). All efforts were made to minimize the number of animals used and their suffering.

### Immunofluorescence

*Mocha* and control (heterozygous littermates) 1-2 months old mice were anaesthetized with 35% chloral hydrate or pentobarbital and perfused through the heart with paraformaldehyde 4%. The dissected brains were fixed overnight in 4% paraformaldehyde in phosphate buffered saline and cut on a vibratome to 30μm thick sections and processed for immunofluorescence as described previously (9). Confocal laser-scanning microscopy was performed using a SP2 confocal microscope (Leica Microsystems, Mannheim, FRG). Images were assembled using Adobe Photoshop (Adobe Systems, San Jose, CA, USA).

### Western Blotting

SDS-PAGE analysis was performed using 4-12% NuPAGE (Invitrogen, France) gradient gels and the manufacturer's buffers and then processed for Western blotting. Blots were quantified using ImageJ (NIH, Maryland) and statistical significance was estimated using Mann-Whitney rank sum tests performed under SigmaStat (SPSS Inc.)

### Electrophysiological recordings

Organotypic hippocampal slices from 6 days old mice were maintained in culture as described previously (40, 41). After 2-3 days, cultures were preincubated 24 hours in serum-free medium and then grown another 3-4 days in fresh serum-free medium containing TeNT (500 ng/ml) or not. Electrophysiological recordings were carried out as described in supplementary methods. For mossy fiber stimulation, the stimulating electrode was positioned at the hilar border of the granule cell layer. For Schaffer collateral stimulation, a cut was made between areas CA3 and CA1 and the electrode was positioned in stratum radiatum ~50-100 μm apart from the recorded cell.

Values are expressed as mean±SEM. Statistical significance was estimated using Mann-Whitney or Wilcoxon rank sum tests, or two-way ANOVA performed under SigmaStat (SPSS Inc.).

**Acknowledgements:**

We are grateful to Lucien Cabanié for production of Cl158.2, Mathilde Chaineau and Agathe Van der Linden for help with *mocha* mice, and Richard Miles for support and critical reading of the manuscript. This work was supported in part by grants from INSERM (Avenir Program), the European Commission ('Signaling and Traffic' STREP 503229), the Association Française contre les Myopathies, the Ministère de la Recherche (ACI-BDP), the Fondation pour la Recherche Médicale, the HFSP (RGY0027/2001-B101) and the Fondation pour la Recherche sur le Cerveau to TG. RR was supported by a postdoctoral fellowship from the Fondation pour la Recherche Médicale, AS by a fellowship from the Swiss National Science Foundation, and LD by a post-doctoral fellowship from the Association pour la Recherche sur le Cancer.

**References**

1. Galli, T. & Haucke, V. (2004) *Sci STKE* **2004,** re19.
2. Takei, K., Mundigl, O., Daniell, L. & De Camilli, P. (1996) *J.Cell.Biol.* **133,** 1237-1250.
3. Sudhof, T. C. (2004) *Annu Rev Neurosci* **27,** 509-547.
4. Faundez, V., Horng, J. T. & Kelly, R. B. (1998) *Cell* **93,** 423-32.
5. Simpson, F., Peden, A. A., Christopoulou, L. & Robinson, M. S. (1997) *J Cell Biol* **137,** 835-45.
6. Dell'Angelica, E. C., Ohno, H., Ooi, C. E., Rabinovich, E., Roche, K. W. & Bonifacino, J. S. (1997) *Embo J* **16,** 917-28.
7. Kamiya, H., Shinozaki, H. & Yamamoto, C. (1996) *J Physiol* **493 (Pt 2),** 447-55.
8. Schoch, S., Deak, F., Konigstorfer, A., Mozhayeva, M., Sara, Y., Sudhof, T. C. & Kavalali, E. T. (2001) *Science* **294,** 1117-22.
9. Muzerelle, A., Alberts, P., Martinez-Arca, S., Jeannequin, O., Lafaye, P., Mazie, J.-C., Galli, T. & Gaspar, P. (2003) *Neuroscience* **122,** 59-75.
10. Martinez-Arca, S., Rudge, R., Vacca, M., Raposo, G., Camonis, J., Proux-Gillardeaux, V., Daviet, L., Formstecher, E., Hamburger, A., Filippini, F., D'Esposito, M. & Galli, T. (2003) *Proc Natl Acad Sci U S A* **100,** 9011-9016.
11. Mallard, F., Tang, B. L., Galli, T., Tenza, D., Saint-Pol, A., Yue, X., Antony, C., Hong, W., Goud, B. & Johannes, L. (2002) *J Cell Biol* **156,** 653-654.
12. Rizzoli, S. O., Bethani, I., Zwilling, D., Wenzel, D., Siddiqui, T. J., Brandhorst, D. & Jahn, R. (2006) *Traffic* **7,** 1163-1176.
13. Zeng, Q., Tran, T. T., Tan, H. X. & Hong, W. (2003) *J Biol Chem* **278,** 23046-54.
14. Sjostrom, P. J., Turrigiano, G. G. & Nelson, S. B. (2003) *Neuron* **39,** 641-54.
15. Woodhall, G., Evans, D. I., Cunningham, M. O. & Jones, R. S. (2001) *J Neurophysiol* **86,** 1644-51.
16. Kantheti, P., Qiao, X., Diaz, M. E., Peden, A. A., Meyer, G. E., Carskadon, S. L., Kapfhamer, D., Sufalko, D., Robinson, M. S., Noebels, J. L. & Burmeister, M. (1998) *Neuron* **21,** 111-22.
17. Vogt, K., Mellor, J., Tong, G. & Nicoll, R. (2000) *Neuron* **26,** 187-96.
18. Faundez, V., Horng, J. T. & Kelly, R. B. (1997) *J Cell Biol* **138,** 505-15.
19. Blagoveshchenskaya, A. D., Hewitt, E. W. & Cutler, D. F. (1999) *Mol.Biol.Cell* **10,** 3979-3990.
20. Salazar, G., Love, R., Werner, E., Doucette, M. M., Cheng, S., Levey, A. & Faundez, V. (2004) *Mol Biol Cell* **15,** 575-87.
21. Hayashi, T., McMahon, H., Yamasaki, S., Binz, T., Hata, Y., Südhof, T. C. & Niemann, H. (1994) *EMBO J.* **13,** 5051-5061.
22. Giraudo, C. G., Eng, W. S., Melia, T. J. & Rothman, J. E. (2006) *Science*.
23. Xu, T., Binz, T., Niemann, H. & Neher, E. (1998) *Nat.Neurosci.* **1,** 192-200.
24. Lonart, G. & Sudhof, T. C. (2000) *J Biol Chem* **275,** 27703-27707.
25. Rosenmund, C. & Stevens, C. F. (1996) *Neuron* **16,** 1197-207.
26. Salazar, G., Craige, B., Styers, M. L., Newell-Litwa, K. A., Doucette, M. M., Wainer, B. H., Falcon-Perez, J. M., Dell'angelica, E. C., Peden, A. A., Werner, E. & Faundez, V. (2006) *Mol Biol Cell*.

27. Salazar, G., Love, R., Styers, M. L., Werner, E., Peden, A., Rodriguez, S., Gearing, M., Wainer, B. H. & Faundez, V. (2004) *J Biol Chem* **279,** 25430-9.
28. Salazar, G., Craige, B., Wainer, B. H., Guo, J., De Camilli, P. & Faundez, V. (2005) *Mol Biol Cell* **16,** 3692-704.
29. Lopantsev, V., Wenzel, H. J., Cole, T. B., Palmiter, R. D. & Schwartzkroin, P. A. (2003) *Neuroscience* **116,** 237-48.
30. Stobrawa, S. M., Breiderhoff, T., Takamori, S., Engel, D., Schweizer, M., Zdebik, A. A., Bosl, M. R., Ruether, K., Jahn, H., Draguhn, A., Jahn, R. & Jentsch, T. J. (2001) *Neuron* **29,** 185-196.
31. Grabner, C. P., Price, S. D., Lysakowski, A., Cahill, A. L. & Fox, A. P. (2006) *Proc Natl Acad Sci U S A*.
32. Sakaba, T., Stein, A., Jahn, R. & Neher, E. (2005) *Science* **309,** 491-4.
33. Goda, Y. & Stevens, C. F. (1994) *Proc Natl Acad Sci U S A* **91,** 12942-6.
34. Li, C., Ullrich, B., Zhang, J. Z., Anderson, R. G., Brose, N. & Sudhof, T. C. (1995) *Nature* **375,** 594-599.
35. Sugita, S., Shin, O. H., Han, W. P., Lao, Y. & Sudhof, T. C. (2002) *Embo J* **21,** 270-280.
36. Rao, S. K., Huynh, C., Proux-Gillardeaux, V., Galli, T. & Andrews, N. W. (2004) *J. Biol. Chem.* **279,** 20471-20479.
37. Sugita, S., Han, W. P., Butz, S., Liu, X. R., FernandezChacon, R., Lao, Y. & Sudhof, T. C. (2001) *Neuron* **30,** 459-473.
38. Hui, E., Bai, J., Wang, P., Sugimori, M., Llinas, R. R. & Chapman, E. R. (2005) *Proc Natl Acad Sci U S A* **102,** 5210-4.
39. Gutierrez, R. & Heinemann, U. (1999) *Brain Res* **815,** 304-16.
40. Stoppini, L., Buchs, P. A. & Muller, D. (1991) *J Neurosci Methods* **37,** 173-82.
41. Musleh, W., Yaghoubi, S. & Baudry, M. (1997) *Brain Res* **770,** 298-301.

**Figure legends**

**Fig.1**

Evoked synaptic transmission in slice cultures prepared from control and *mocha* mice.

(a) Postsynaptic currents evoked in CA3 pyramidal cells by MF stimulation. 10 superimposed sample traces are shown for maximal stimulation intensity. Top: control cultures. Bottom: cultures treated for at least 72 hours with TeNT (500 ng/ml). After TeNT treatment unreliable, asynchronous responses (marked with a tick) can be detected in control (+/-) but not mocha (-/-) cultures. EPSCs evoked by Schaffer collateral stimulation in CA1 cells are also completely abolished by prior incubation with TeNT. (b) Stimulation arrangement: mossy fibers (MF) were stimulated with a patch pipette located at the hilar border of the granule cell layer (MF stim.). Associational/commissural fibers (A/C) were stimulated in st. oriens. Responses to either stimulation were distinguishable by their sensitivity to DCG-IV (1µM). (c) Comparison of responses (integrated over 200ms of the stimulus artifact) evoked by increasing MF stimulations in control vs. mocha cultures or by SC stimulation in CA1 cells. Left: in control cultures, no significant difference could be detected between control (filled circles) and mocha CA3 cells (open circles, n=4 and 7 cells, respectively; Two way ANOVA p>0.05). Right: after treatment with TeNT, responses were evoked in control CA3 cells (filled triangles) but not in mocha CA3 (open triangles) or control CA1 cells (filled squares, n=7, 11 and 5 cells, respectively; Two way ANOVA, $p<0.02$ for control CA3 vs. mocha CA3 or control CA1 and $p>0.05$ for mocha CA3 vs. control CA1).

**Fig. 2**

TI-VAMP is absent from MF terminals and accumulates in the cell body of *mocha* neurons.

50 µm vibratome sections from adult control (+/-) and *mocha* (-/-) mice were labeled with DAPI (blue) and monoclonal antibodies against TI-VAMP or Syb2, polyclonal antibodies against Syp or VAMP4.

(a) TI-VAMP is lost from the terminals of granule cells (mossy synapses) innervating CA3 pyramidal cells in *mocha* brain. A dense signal for Syp (green) is detected throughout the CA3 region in stratum oriens (so), stratum pyramidale (sp) and stratum radiatum (sr) in control (+/-) hippocampus. The strong immunofluorescence presented in the stratum lucidum (sluc), correspond to MF terminals innervating CA3 pyramidal dendrites. Lower panels are magnification of the boxed windows. In control (+/-) mice, confocal images shows that TI-VAMP (red) immunoreactivity colocalizes in mossy synapses with Syp puncta in sluc. However in mocha (-/-) mice, TI-VAMP is not detected in presynaptic terminals (sluc) of granule cells, while a faint immunofluorescence is detected in pyramidal cell soma (boxed window in a and c). (b). Syb2 localization is unaffected in the mocha hippocampus. As for Syp immunoreactivity, Syb2 is detected in control and *mocha* mice throughout so, sp, sluc and sr. (c, d). TI-VAMP is mislocalized in the granule cells of mocha mice. In control mice, TIVAMP is present in MF both in the lucidum (c) and in the hilus of the dentate gyrus (d). In mocha mice, the strong labeling of the lucidum and hilus in the dentate gyrus is not present. Double labeling of TIVAMP and VAMP4 shows that the remaining TIVAMP colocalizes with VAMP4 in the cell bodies of pyramidal (sp) and granule cells (stratum granulosum, sg, boxed windows). Bar: 100 µm.

**Fig. 3**

Ca-independent, quantal release at excitatory synapses on CA3 cells in control and *mocha* cultured slices.

(a) Representative traces of miniature EPSCs (mEPSCs) recorded in CA3 pyramidal cells from control (+/-) and *mocha* (-/-) cultures, treated or not with TeNT. (b) Averaged mEPSCs (~100) detected from the above recordings. Black traces: control. Blue traces: after TeNT treatment. No difference in either their rate of onset or decay was apparent. (c) Left: average amplitude of mEPSCs recorded in all four conditions. No significant difference was observed (n=7, 9, 11 and 8 cells, respectively. p>0.05). Right: cumulative amplitude histograms from the same four datasets. The distributions were not significantly different (Kolmogorov-Smirnov test, p>0.05). (d) Mean frequencies of mEPSCs were significantly different between control and TeNT-treated cultures in both control and mocha cultures (p<0.005 and p<0.05 respectively). mEPSC frequency was also different in control vs. mocha culture in the absence of TeNT (p<0.05). (e) *Mocha* culture slices treated with TeNT for 72 h were fixed and labeled with antibodies against

19

SNAP25 (red), TeNT (green), and DAPI (blue). The whole surface of the explant can be visualized either by DAPI (nucleus) or by SNAP25 (neuronal plasma membrane). Note that the TeNT staining is uniformly distributed confirming the extended penetration of the toxin. Scale bar: 200 μm. (f) Culture slices used in electrophysiological recordings were lysed and analyzed by Western blotting with antibodies against AP3δ, Syb2, and actin (as a loading control). 72 h treatment with TeNT resulted in efficient cleavage of Syb2 although quantification of the remaining Syb2 revealed a two-fold increase in TeNT-resistant Syb2 in *mocha* slices (mean ± SEM: control 17.83 ± 2.35 n=6, mocha 37.58 ± 6.31 n=6, p<0.015 (Mann-Whitney rank sum test)). (g) Mocha cultured slices treated with or without TeNT for 72 h were fixed and labeled with antibodies against Synaptophysin (green), Synaptobrevin2 (red), and DAPI (blue). Note that the remaining Syb2 labeling after TeNT treatment is very faint. The rare remaining Syb2 puncta are mainly synaptic. Bar: 50 μm.

**Fig. 4**

Differential recruitment of readily releasable vesicles by hypertonic sucrose applications in control vs. *mocha* cultured slices.

(a) Sample traces from individual recordings of mEPSCs recorded in the presence of tetrodotoxin and an external Ca/Mg ratio of 0.1 before and during application of a 500 mOsm sucrose solution in the presence of a 2.5 ml/min normal ACSF perfusion. High sucrose was applied by pressure injection through a patch pipette to the proximal dendrites of the recorded cell in st. lucidum. Low sucrose: ~0.25 psi. High sucrose: 1.5 psi. (b) Average frequency of mEPSCs recorded in all conditions. The sensitivity of mEPSC frequency to sucrose was significantly different in control vs. mocha cultures (n=4 and 6 cells, respectively; Two way ANOVA p<0.01). After treatment with TeNT, sucrose application failed to increase mEPSC frequency (n=9 and 7 cells, respectively). (c) Frequencies were normalized to control values, showing a greater relative increase in *mocha* (open bars) compared to control slices (filled bars) upon low-pressure (Mann-Whitney p<0.05) but not high-pressure application of sucrose (p=0.5).

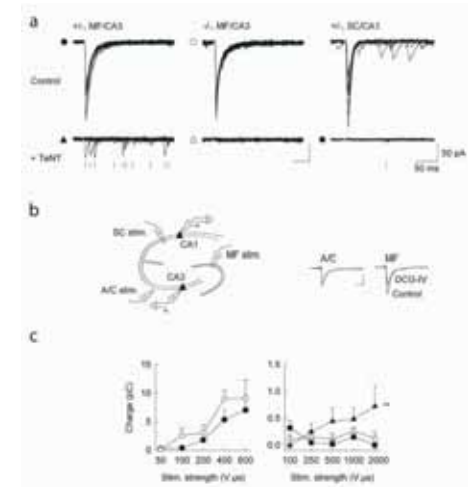

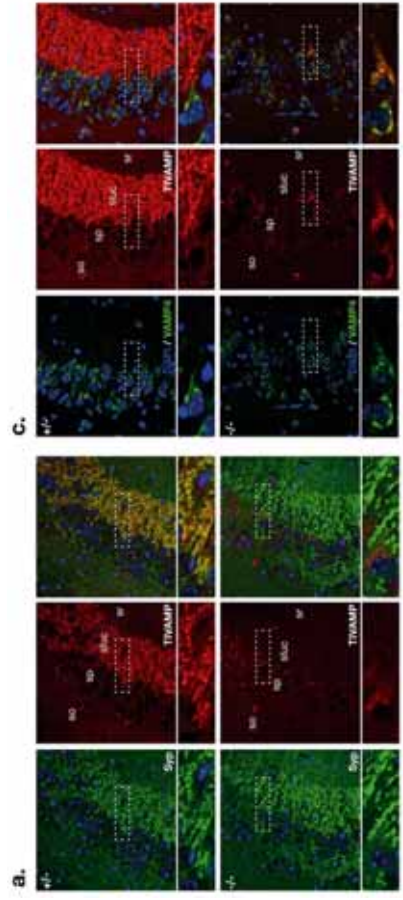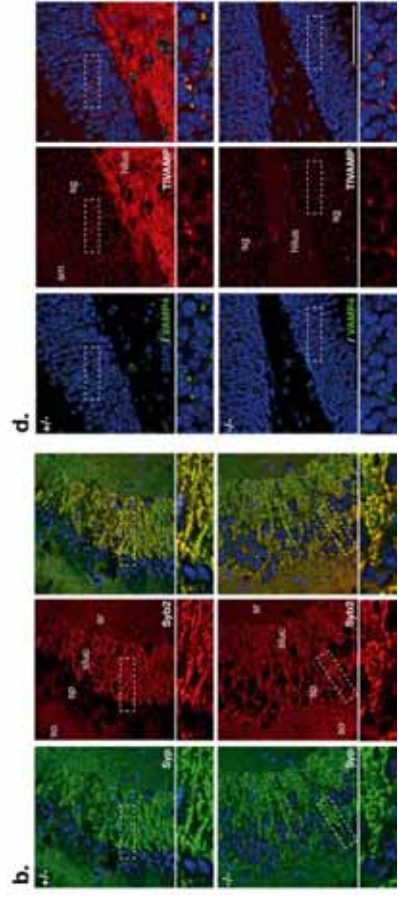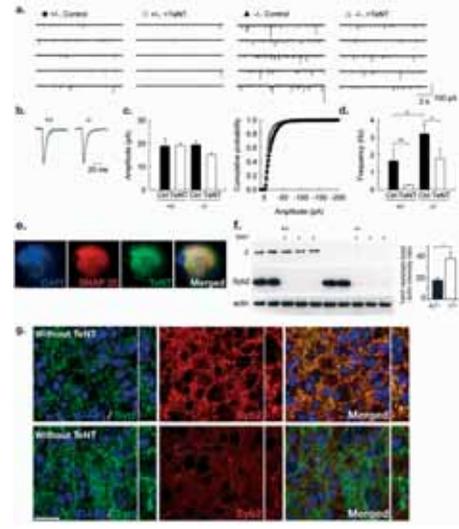

## Supplemental Results

*Late onset of AP-3 expression explains normal* mocha *brain development*

Our previous studies demonstrated that TI-VAMP is critical for neurite outgrowth (1, 2). Although brain development of the *mocha* mice seemed normal (3), such a defect may impair hippocampal connectivity and therefore complicate the interpretation of our present results. We thus examined TI-VAMP localization and neuritogenesis in control and *mocha* neurons (Fig. S3a, b). Both neurite outgrowth and TI-VAMP localization to the growth cone were normal in hippocampal neurons cultured 3div from *mocha* mice (Fig. S3a, b), demonstrating that TI-VAMP is transported to the growth cone in an AP3-independent manner. The lack of function of AP3 in neuritogenesis is further supported by the developmental regulation of its expression. In neurons growing in culture, AP3-δ expression was greatly increased at 7 days in vitro and reached a peak at ~14 days (Fig. S3c). These results are thus consistent both with the essential role for TI-VAMP in neurite outgrowth and with the fact that brain development appears normal in *mocha* mice, suggesting a role of AP3 in SV recycling rather than neurite outgrowth.

## Supplemental Material and Methods

### Antibodies and Reagents

The mouse monoclonal antibody (clone 158.2; (4) and rabbit polyclonal antibody directed against TI-VAMP (TG18; (5), the rabbit polyclonal antibodies anti-synaptophysin (MC1; (6)), anti-VAMP4 (TG19; (7)) and anti-SNAP-25 (MC9; (8)) have been described previously. Mouse monoclonal antibodies against Syb2 (clone 69.1), Rab3a (clone 42.2) and Synaptotagmin 1 (clone 41.1) were generous gifts from R. Jahn (Max Planck Institute, Goettingen, FRG). Monoclonal antibody against TeNT has been described previously (9). Monoclonal anti-AP3δ antibody (SA4) was from the Developmental Studies Hybridoma Bank (University of Iowa, Iowa City, Iowa). Chicken polyclonal antibody against MAP2 (ab5392) was from AbCam (Milton Road, Cambridge, UK). Affinity-purified horseradish peroxidase-, Cy3- and Cy5-coupled biotin-coupled goat anti-rabbit and anti-mouse immunoglobulins were from Jackson ImmunoResearch (West Grove, PA, USA). Alexa 488- and Alexa 594-coupled goat anti-rabbit and anti-mouse immunoglobulins and Alexa 568-coupled phalloidin were from Molecular Probes (Carlsbad, CA). Tetanus neurotoxin was isolated and characterized as previously described (10).

### Neuronal Cell Culture and Immunocytochemistry

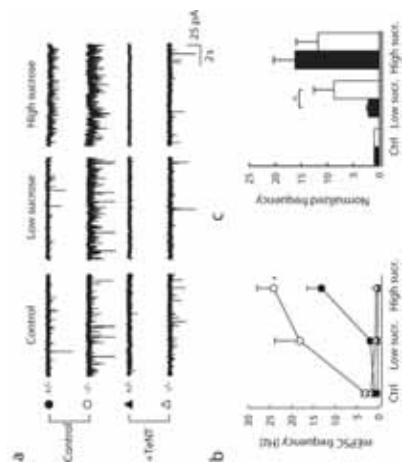

Hippocampal neurons from newborn (P0-P1) mice were prepared as described previously (11) and grown on poly-L-Lysine-coated (Sigma-Aldrich, France) 14mm coverslips at a density of 25,000-50,000 per coverslip in Neurobasal media supplemented with B27 (GIBCO, Invitrogen France) and Cytosine β-D-arabinofuranoside (Sigma-Aldrich, France). Granule cells were obtained from dissected dentate gyrus of postnatal mice (P5) (12, 13) and grown on poly-Ornithine-coated (Sigma-Aldrich, France) per 14mm coverslips at a density of 250,000 per coverslip in Neurobasal media supplemented with B27 (GIBCO, Invitrogen France), 2mM Glutamine, 25% horse serum for the first 5div, 10% thereafter. Neurons in culture were fixed with 4% paraformaldehyde/ 4% sucrose and processed for immunofluorescence microscopy as described previously (5).

**Electrophysiological recordings**

For recording, cultures were transferred to a submerged chamber mounted on an upright microscope and superfused at a rate of ~2.5 ml/min with ACSF containing (in mM): NaCl, 124, $NaHCO_3$, 26.2, D-glucose, 11, KCl, 2.5, $NaH_2PO_4$, 1, $CaCl_2$, 4 and $MgCl_2$, 4. Whole-cell recordings from CA3 or CA1 pyramidal cells were made using patch electrodes (3-5 MΩ resistance) made from borosilicate glass capillaries (Hilgenberg, Germany) and filled with internal solution containing (in mM): $CsCH_3SO_3$, 110, CsCl, 20, HEPES, 10, EGTA, 10, MgATP, 4, $Na_3GTP$, 0.4, $MgCl_2$, 1.8.

Postsynaptic currents were evoked at 0.1 Hz using extracellular stimulation through a borosilicate microelectrode filled with ACSF and recorded from cells held at -60 mV using an Axopatch 1D amplifier and digitized at 20 kHz using Clampex software of the pClamp 9 suite (Axon Instruments, USA). PSC amplitude and access resistance were monitored online. PSCs were then measured offline using Clampex software. Currents were integrated over a 200 ms time window set immediately after the stimulus artifact ($C_{200}$). In order to correct evoked PSC charge for spontaneous PSCs, baseline currents were also integrated from a 50 ms pre-stimulus period and the corrected charge was calculated as $C_{corr} = C_{200} - 4 \times C_{50}$.

Miniature EPSCs were recorded in modified ACSF in which $CaCl_2$ and $MgCl_2$ concentrations were adjusted to 0.5 and 4.5 mM, respectively, in the presence of tetrodotoxin (TTX, 2 µM, Latoxan, France) and bicuculline methochloride (20 µM, Fisher Bioblock Scientific, France). For focal application of a hypertonic solution, a second patch pipette filled with 0.5M sucrose in ACSF was positioned in st. lucidum ~40-50 µm apart from the recorded cell and connected to a PicoSpritzer device (General Valve, Fairfield, NJ, USA). Miniature EPSCs were detected using Detectivent software written under LabView (kindly provided by N. Ankri, INSERM U641, Marseille, France).

In some experiments, acute hippocampal slices were prepared from 12-14 days old pups as described (14). Slices were kept ~1 hour in normal ACSF equilibrated with a 95% $O_2$ / 5% $CO_2$ mixture. Half of the slices were then exposed to 10µg/ml brefeldin A (Sigma, France) for an additional 2 hours whereas the other half was kept in ACSF until recordings were started.

**Electron Microscopy**

Neurons from control or *mocha* mice were fixed with 2% (w/v) PFA or with a mixture of 2% (w/v) PFA and 0.2% (w/v) glutaraldehyde in 0.1 M Phosphate Buffer (PB) pH 7.4. Cells were processed for ultracryomicrotomy as described (15). Ultrathin sections were prepared with an Ultracut FCS ultracryomicrotome (Leica, Wien, Austria), and single immunogold labeled with antibodies and protein A coupled to 10 or 15 nm gold (PAG 10 and PAG 15). Sections were observed and photographed under a Philips CM120 Electron Microscope (FEI Company, Eindoven, The Netherlands). Digital acquisitions were made with a numeric Keen View camera (Soft Imaging System, Germany).

**Neurite outgrowth measurements**

Neurons were fixed at 1 div, labeled with Alexa-568-coupled phalloidin, and imaged with a MicroMax CCD camera (Princeton Instruments, Princeton, NJ). Neurite length was measured using Metamorph software (Princeton Instruments).

**Supplemental Figures**

**Supplemental Figure 1: TIVAMP and synaptic proteins localization in the whole hippocampus of control and *mocha* mice.**

a. AP3δ (red) is expressed in the hilus in the dentate gyrus (DG) and in stratum lucidum in CA3 region of control (+/-) mice but not in mocha (-/-) mice. b. TIVAMP (red) is mainly expressed by granule cells in presynaptic terminals of the hilus in the dentate gyrus and in stratum lucidum in CA3 region and faint or absent in pyramidal terminals of the CA1 region in control (+/-) mice. TIVAMP expression shows a strikingly different pattern in mocha (-/-) mice. Syp immunoreactivity (green) is similar in both mice. The nuclei are labeled with DAPI (blue). Scale bar: 200 µm. The boxed region corresponds to the field seen in figure 1 at high magnification. c. Distribution patterns of synaptic vesicle proteins were assayed in control and mocha mice. No major default was noticed for Syb2, and Rab3a (red) immunoreactivities between the two populations of mice.

**Supplemental Figure 2: TI-VAMP concentrates in synapses in cultured granule cells and accumulates in the Golgi of mature *mocha* pyramidal neurons.**

(a) Granule cells (21div) from control mice were colabeled with antibodies against TI-VAMP, Syp, and MAP2. TI-VAMP colocalizes with Syp (67.20+/-6.65% of Syp positive punctae are also TI-VAMP positive, 10 cells, 1282 synapses), suggesting that it is present in a large subset of synaptic terminals. Post-synaptic neurons contacted by Syp+ terminals (neuron, n) were identified on the basis of the MAP2 staining to exclude astroglial (glial, g) staining of TI-VAMP. (b) Hippocampal neurons (14 div) cultured from control (+/-) or *mocha* (-/-) mice were colabeled with antibodies against TI-VAMP and antibodies against VAMP4. In *mocha* neurons TI-VAMP accumulates in the Golgi where it colocalizes with VAMP4. Bar: 10μm. (c) Ultrathin cryosections of mature cultured hippocampal neurons (14 div) from control or *mocha* mice were immunogold labeled with polyclonal antibodies against TI-VAMP. In control (+/-) neurons, TI-VAMP is present on vesicles and tubules throughout the soma and occasionally present at the plasma membrane, while in *mocha* (-/-) neurons TI-VAMP labeling is mainly found associated with the Golgi apparatus (GA).

**Supplemental Figure 3: Neurite outgrowth and TI-VAMP localization to the growth cone are unaffected in mocha neurons.** (a) Immature hippocampal neurons (3 div) cultured from control (+/-) or *mocha* (-/-) newborn mice were labeled for TI-VAMP and analyzed by confocal microscopy. In both control and *mocha* neurons, TI-VAMP labeling is present in the growth cone. Bar: 10μm (b) Hippocampal neurons from control (+/-) and *mocha* (-/-) mice were grown for 1 div and their neurite length was measured. No difference in neurite outgrowth was observed between control and *mocha* neurons (Bars represent SEM; n=100 neurons from two independent experiments). (c) Hippocampal neurons growing in culture were lysed at different stages in development and TI-VAMP and AP3 expression were analyzed by Western blotting. An equal quantity of protein was loaded in each lane. Note that both TI-VAMP and AP3 expression are upregulated during development.

**Supplemental Figure 4: Increased quantal release at MF terminals after acute disruption of AP3 by Brefeldin A.**

(a) Sample traces from individual recordings of mEPSCs recorded from CA3 pyramidal cells in acute hippocampal slices (P10-12). Prior to recording, slices were incubated for > 2 hours in either normal ACSF or ACSF containing 10μM Brefeldin A. In Brefeldin-treated slices, mEPSC frequency was increased by ~127 % as compared to control ($2.59 \pm 0.33$ vs. $1.14 \pm 0.25$ Hz, n=14 and 12 cells, respectively, $p<0.005$). (b) Scaled and averaged mEPSCs (~100) detected from the recordings shown in (a). Black traces: control. Blue traces: Brefeldin A. No difference in either their rate of onset or decay was apparent. (c) Summary data from all recordings. (d) Ratio of TI-VAMP:Syp fluorescence intensity in control and BFA-treated slices in the hilus and str.lucidum (mean+/-sem, n=14 slices).

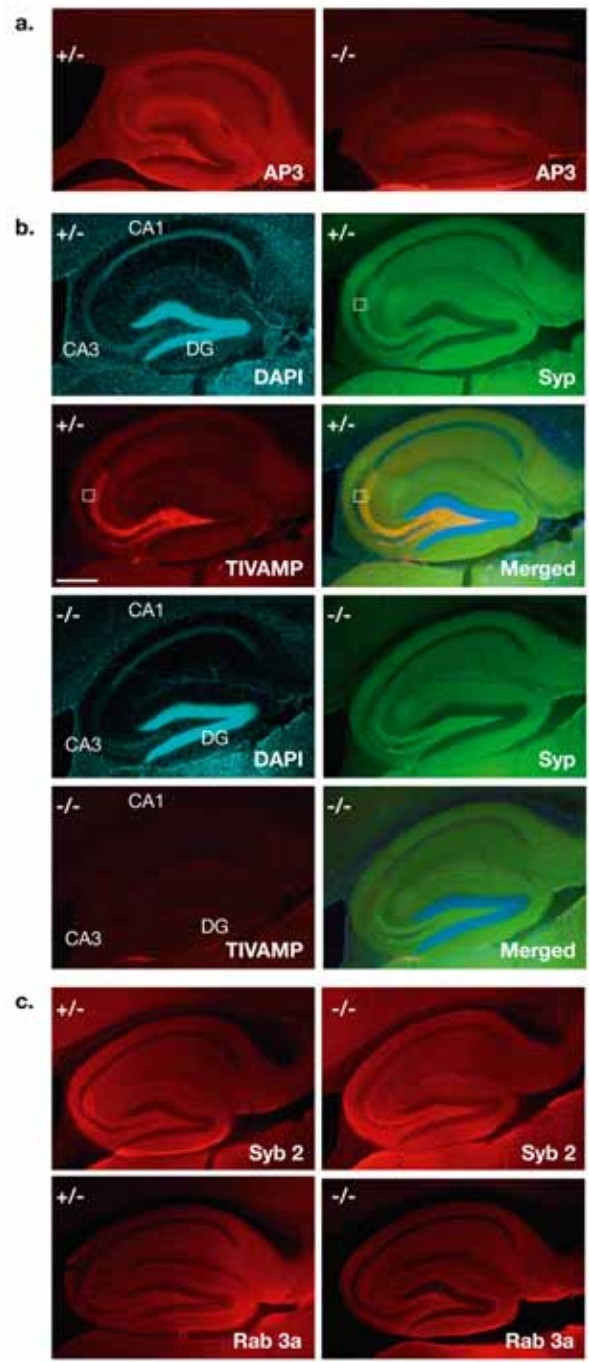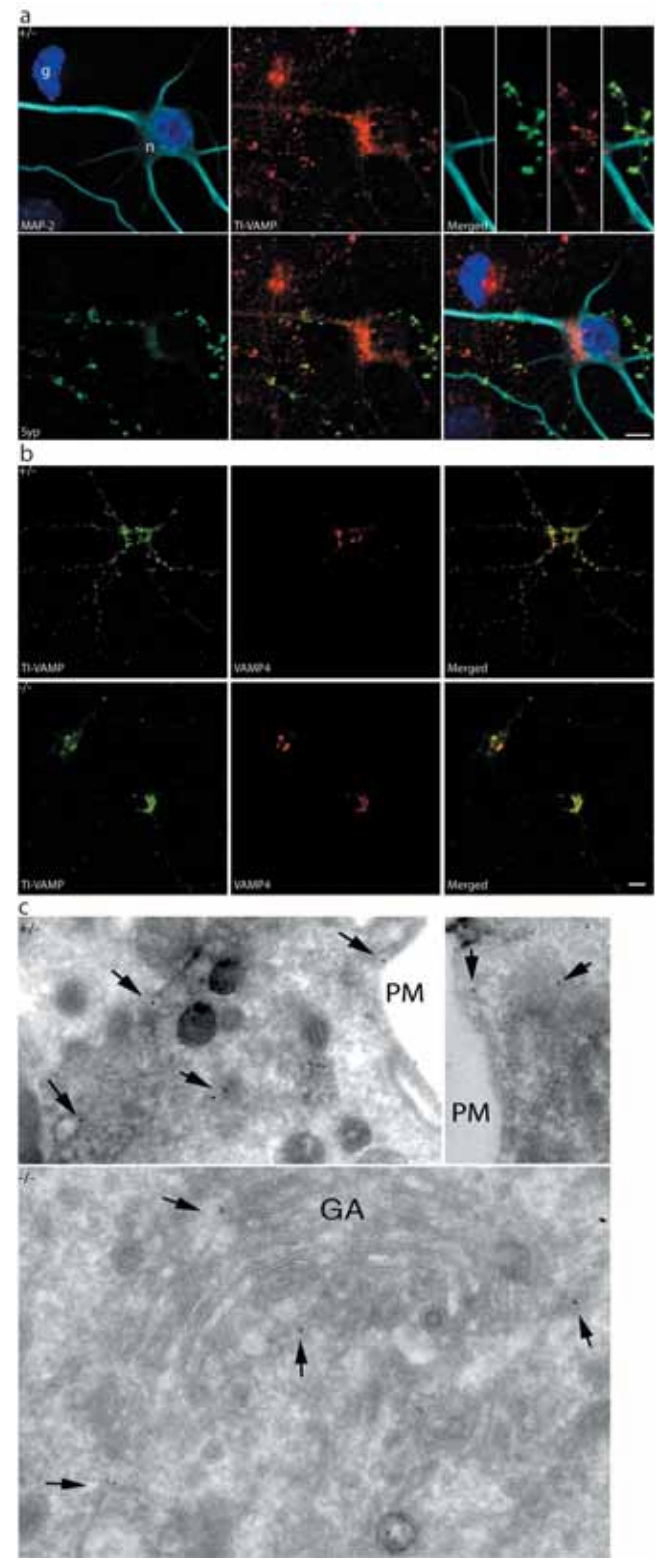

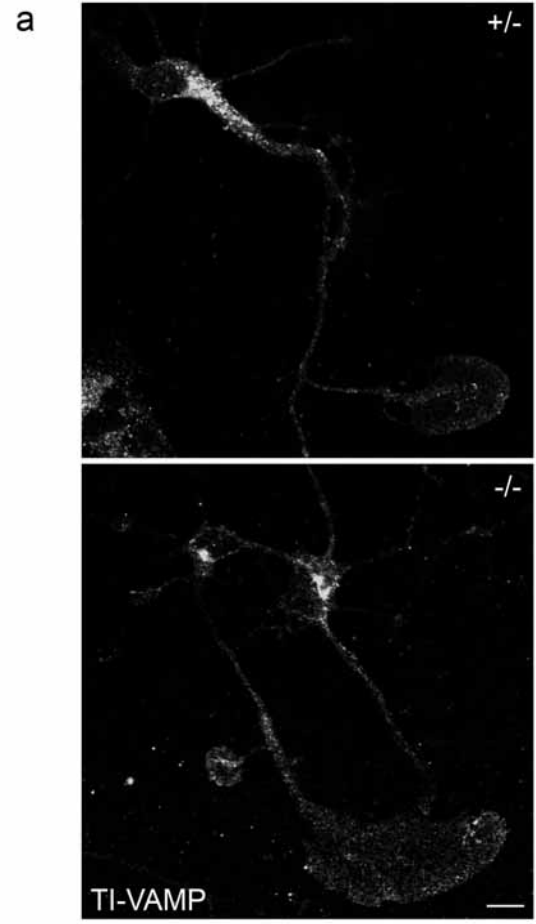
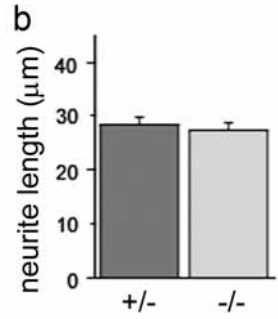
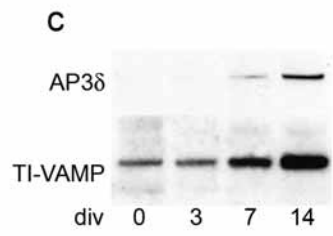
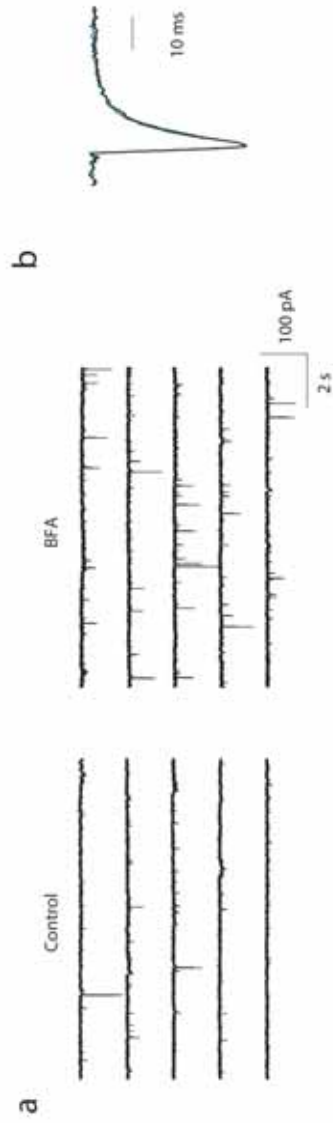
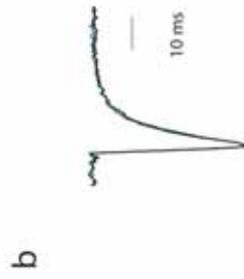
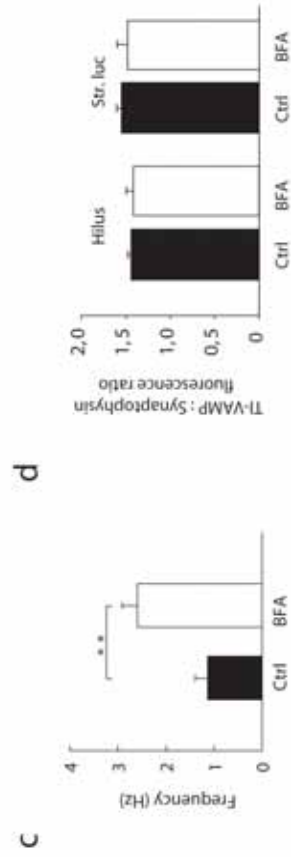